# Unveiling the Impact of Macroeconomic Policies: A Double Machine Learning Approach to Analyzing Interest Rate Effects on Financial Markets

**Anoop Kumar [1], Suresh Dodda[2], Navin Kamuni [3], Rajeev Kumar Arora[4]**

[1]Anoop.kumar.2612@gmail.com, [2] sureshr.dodda@gmail.com, [3]navin.kamuni@gmail.com, [4]rajeev04.study@gmail.com

[1] IIT Roorkee, India
[2] IT Department, Eudoxia Research center, USA
[3]AI-ML, BITS Pilani WILP, USA
[4]Ph.D Himalayan University, India

*Abstract*— **This study examines the effects of macroeconomic policies on financial markets using a novel approach that combines Machine Learning (ML) techniques and causal inference. It focuses on the effect of interest rate changes made by the US Federal Reserve System (FRS) on the returns of fixed income and equity funds between January 1986 and December 2021. The analysis makes a distinction between actively and passively managed funds, hypothesizing that the latter are less susceptible to changes in interest rates. The study contrasts gradient boosting and linear regression models using the Double Machine Learning (DML) framework, which supports a variety of statistical learning techniques. Results indicate that gradient boosting is a useful tool for predicting fund returns; for example, a 1% increase in interest rates causes an actively managed fund's return to decrease by -11.97%. This understanding of the relationship between interest rates and fund performance provides opportunities for additional research and insightful, data-driven advice for fund managers and investors.**

## I. Introduction

Interpreting the impact of interest rates, where a hypothetical 1% increase in rates implies an 11% decline in returns from actively managed funds, is central to the study. This substantial impact is consistent with theory, but given the variety of factors affecting market performance, further investigation is required. The intricate dynamics of the financial market are exemplified by the quick market response to interest rate changes, as opposed to the delayed responses in other economic sectors. The flexibility of the Double Machine Learning (DML) framework, indicating the possibility of its wider use in financial research by various studies such as [1]–[3]. However, the studies acknowledges certain limits, such as the difficulties presented by complex data and the requirement for advanced modelling methods in order to precisely represent the underlying economic events.

This study examines the effectiveness of the DML framework for evaluating the relationship between fund returns and the growth in interest rates of the US Federal Reserve System (FRS), especially for actively and passively managed funds between January 1986 and December 2021. Approximately 7,000 funds are involved in this study, which uses gradient boosting and linear regression models to demonstrate the intricacy of the financial sector. Because of its known impact on market dynamics, DML is proposed as a novel method for evaluating average treatment impacts, with a particular emphasis on the interest rate of the FRS. One of the main questions in financial analysis is whether DML is feasible and can accurately assess causal effects on fund returns. Preliminary results show that DML aligns well with the intricacies of financial data, but the method needs to be handled carefully because financial markets are complex. Furthermore, gradient boosting exhibits strong predictive ability, suggesting its possible use in financial DML. Therefore, the empirical research conducted for the study shows a strong negative correlation, supported by a highly precise gradient boosting model, between interest rate growth and fund returns for actively managed funds. This relationship is consistent with macroeconomic models that postulate this kind of causal relationship, emphasizing how sensitive fund returns are to changes in monetary policy. However, the findings for passively managed funds were inconsistent, indicating that more research is required to fully comprehend the nuances of this market.

The paper is as follows; related studies are shown in the following section. Data and model analysis are offered along with the material and methods in Section III. The experimental results are shown in Section IV. The discussion is offered in Section V, and the study is concluded with some conclusions and ideas for future work in Section VI.

## II. STATE OF THE ART

The study of financial return forecasting has attracted a lot of attention from academics and industry professionals in an effort to improve investment strategies by using predictive analytics. Prominent research has highlighted the possible advantages of these projections in directing investment choices ([4], [5]), but they frequently fail to clarify the fundamental reasons for market fluctuations. In order to demystify the mechanisms behind volatility trends in financial markets and to facilitate more informed, data-driven decision-making, this study argues for the incorporation of causal inference into financial analysis. The foundational techniques for identifying causal connections have been supplied by traditional econometric methodologies, such as Granger[1] causality. But often these methods aren't effective enough to prove true causation, especially when dealing with the kind of intricate, massive datasets that the finance industry uses. Recent developments have combined statistical learning with traditional causal inference to solve these issues such as [6], [7], providing more sophisticated analytical methods that can handle the complexities of financial data while still reaching conceptually sound results. These innovative approaches are being used more and more in many socio-economic fields, but they are still not widely used in finance. In fact, there are very few studies using real data to study financial issues such as [8]–[10].

Therefore, this highlights a need for more research, especially to help managers and investors improve their strategies with causally informed insights. The analysis of actively and passively managed funds is the main focus of this study, which uses information from almost 7,000 funds. Given their varying sensitivity to market dynamics and their collective representation of wider market trends, the differentiation between fund kinds is relevant. This study incorporates gradient boosting and linear regression models into the DML framework. By making it easier to estimate average treatment effects, the DML framework allows for a more sophisticated understanding of the causal effects on financial markets. This analysis centers on the role of FRS interest rates, a factor that has been repeatedly emphasized in the media and academia as a major contributor to market volatility. This supports the idea that sophisticated statistical learning techniques can improve the accuracy of financial projections.

## III. MATERIALS AND METHODS

In financial analysis, it is critical to distinguish between prediction and causality. While causation explores the causes that lead to these outcomes, prediction concentrates on finding patterns that point to future events. The definition by [11] is used in this study, with a focus on the causal link where variable X affects the value of Y. In the financial markets, where investor behavior frequently reacts to macroeconomic signals like FRS policies, this viewpoint is especially pertinent. With no feedback loops, Directed Acyclic Graphs (DAGs) provide a visual depiction of causal assumptions by indicating the direction of influence among variables [12], [13]. The DAG structure is put to the test by the feedback that financial markets frequently display, which allows investor behavior to impact FRS decisions. That being said, the study concentrates on the one-way impact of interest rate policies set by the US FRS on market behaviors. Although we use different approaches and different assumptions, causal inference and prediction have comparable goals. Forecasting outcomes is the goal of prediction tasks, which do not take into account the causal relationships between variables.

ML techniques are mostly focused on prediction jobs because of their capacity to manage enormous datasets and intricate data structures [14]–[23]. Nevertheless, observational data present a problem for traditional causal inference methods since they sometimes rest on non-testable assumptions. This has prompted a search for techniques that can balance the rigors of causal analysis with the complexity of financial data. Therefore, the application of DML is highlighted in the paper. DML combines the predictive power of ML with the meticulous examination of conventional causal inference. By enabling any ML technique to estimate functions and so lowering bias and variation in parameter estimations, DML tackles the problems of confounding and high-dimensional data. This approach makes it easier to estimate the average causal impact of factors on financial outcomes, such as changes in interest rates at the FRS. To apply DML, the analysis is divided into two categories: $\beta$-tasks and $y$-tasks. The $\beta$-tasks are concerned with estimating causality, while the $y$-tasks concentrate on prediction. Partialling out the effects of control variables through this separation allows confounding to be addressed, as ML models trained on a portion of the data can predict outcomes and treatments on the remaining data. Additionally, as ways to improve the predictive and causal analysis, the study examines gradient boosting and autoregressive models. Given the historical values of variables and their correlations, autoregressive models—such as Vector Autoregressive (VAR) models—are well-established instruments for financial return forecasting. On the other hand, gradient boosting—which increases prediction accuracy by repeatedly leveraging weak learners—is emphasized for its efficacy in handling non-linear interactions that are typical in financial markets. This research offers a framework for analyzing the effects of macroeconomic policies on financial markets, contributing to the continuing discussion on causality in finance through the application of innovative methods such as DML when paired with conventional econometric and ML techniques.

### A. Data Analysis

Using data from 14,816 funds, this study explores how macroeconomic policies affect U.S. traded funds. A selection procedure was required due to Bloomberg terminal limitations, which show a maximum of 5,000 funds. In order to ensure that every fund that was listed was taken into account, criteria were created in order to reduce the dataset to a manageable level. The final dataset included 1,948 passively managed funds and 5,000 actively managed funds. Metadata was recorded for the funds, including their names, asset classes, start dates, and assets under management. The data, which was sourced using the Python library covered the period from December 1985 to December

---

[1] http://var.scholarpedia.org/article/Granger_causality

2021, was based on tickers that were retrieved from Bloomberg[2] (refer to Table I). This time frame was selected to capture the changes in fund behaviour both before and after major financial crises, as well as the introduction of new fund categories. The Federal Reserve Economic Data (FRED) was the primary source of information for the macroeconomic variables used in the study, which included the money supply, FRS interest rates, unemployment rates, inflation, and GDP growth rate. In order to align with the treatment variable's update rate and most of the FRED data, the data frequency was set to monthly. Because there aren't many quarterly variables, the risk of adding noise into the monthly data adjustment process is modest. Quarterly data were converted to monthly using either repetition or interpolation. While passively managed funds produced a 432 x 1,948 matrix with little data loss during collection, the actively managed funds dataset created a 432 x 4,976 matrix. One major problem was missing data, which resulted in fewer observations and possible discrepancies when modelling fund returns ($y$), particularly for funds with shorter lifespans. Although the number of funds expanded after 1996 and then again around 2008, the data for the early years was sparse, therefore an effective strategy was required to model the funds consistently and successfully over the whole period.

TABLE I
AVAILABLE FILTERING PARAMETERS AND ACTIVELY MANAGED FUNDS

| Filtering Criteria | Number of Funds |
| --- | --- |
| Market Status: Active | 426,366 |
| Fund Primary Share Class = Yes | 139,777 |
| Country/Territory of Domicile: United States | 14,816 |
| Inception Date >= 1/1/1985 | 14,007 |
| Fund Asset Class Focus: Fixed Income, Equity | 10,996 |
| Fund Actively Managed = Yes | 5,645 |
| Parent Company Name | 5,645 |
| Fund Industry Focus | 5,645 |
| Fund Total Assets (mil) >= 20M | 4,976 |

*1) Data Preprocessing*

Using the average value of Y to represent the various fund performances, an initial correlation analysis was carried out to identify correlations between the X variables and between X and Y prior to the use of ML techniques as shown in Fig. 1 and Fig. 2. Significantly, real Gross Domestic Product (GDP) and inflation showed higher than predicted relationships, which raised questions regarding possible confounding effects in causal analysis. The data were detrended to first differences in order to address the strong inter-variable correlation and non-stationarity. This allowed the focus to shift from absolute values to growth rates. The Augmented Dickey-Fuller (ADF) test confirmed that this strategy, which was in line with [24] technique, successfully decreased variable correlation and made the majority of variables stationary. After detrending, only the natural rate of unemployment remained non-stationary, and it was thus removed from the analysis. The study decided to use VAR models to find the best lag time for the variables instead of using them directly for causality assessment. Based on AIC scores, the study settled on a seven-month lag. As a result of this choice, funds with inadequate historical data were excluded, guaranteeing uniformity throughout the time series study. When switching to panel data, the original 432 x 4,987 matrix of the Time Series Cross-Section (TSCS) was changed to a more manageable 906,205 x 12 structure in order to accommodate the large dataset. This modification made it necessary to include fixed effects in the model in order to take into consideration the non-random character of some variables, such as fund tickers. In order to incorporate fixed effects and add the average of each variable for particular groups without unduly increasing the number of variables, means-encoding was chosen. The high number of unique funds in the dataset needed to be managed, and this method's computing efficiency and preservation of information made it superior to previous encoding strategies. The model's ability to capture the complex interactions seen in the data was improved by the addition of fixed effects and lag variables. This helped to reduce the bias caused by missing variables and increased the validity of the causal conclusions derived from the research.

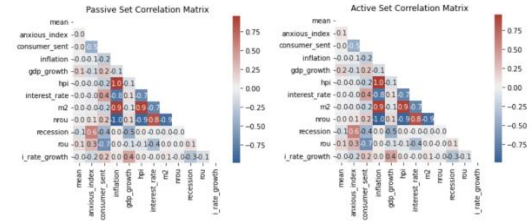

Fig. 1. The correlation matrix of the active and passive datasets

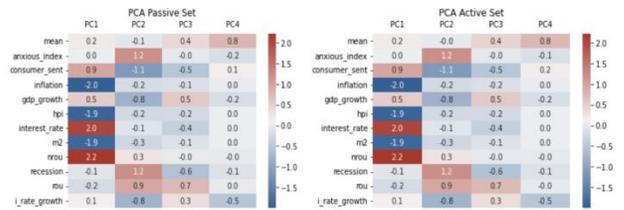

Fig. 2. Principal component analysis of the correlation matrix

*B. Model Analysis*

This study used Python's scikit-learn module to enable the use of statistical learning techniques for financial data analysis, particularly gradient boosting and linear regression. Known for its ease of use and interpretability, linear regression provided a baseline against which other models could be compared, enabling a preliminary evaluation of the predictive power of the model. Even though linear regression is widely available, it has limits when it comes to capturing the variance of the model. This highlights the need for more advanced techniques in order to meet the requirements for DML. The more successful option turned out to be gradient boosting, which demonstrated better modelling results for both active and passive fund datasets. However, the procedure of fine-tuning parameters was more complex using this technique. The study used the *xgb*[3] module from the scikit-learn library and carefully adjusted the model's parameters. The study carefully adjusted the model's parameters using the *xgb* module of the scikit-learn toolkit. Because of processing limitations, a two-fold cross-validation procedure was applied to the parameters that were initially considered for a five-fold process. This phase demonstrated gradient boosting's high computing cost, particularly when controlling for variables like tree depth or number. As demonstrated by better Mean Squared Error (MSE) and $R^2$ scores, parameter adjustment was essential to improving the model's predictive accuracy. This painstaking tuning sought to strike a compromise between maximising $R^2$ and minimising MSE, reflecting the competing goals of prediction accuracy and dependability. Despite being arbitrarily due to hardware constraints, the tuning parameters selection played a crucial role in showcasing gradient boosting's potential for financial data processing. The active funds dataset grew substantially after lagged variables and fixed effects were added, adding another level of complexity to the datasets and highlighting the difficulties in handling high-dimensional data in computer modelling. Although not all-inclusive, the ultimate choice of tuning parameters provided some first understanding of the best configurations for gradient boosting given the limitations of the research. After establishing the optimal settings for gradient boosting, the study proceeded to smoothly incorporate these techniques into the DML architecture. The DoubleML[4] library made this process easier by automating tasks like refitting, cross-validation, and sample splitting, which simplified the use of DML. Researchers wishing to investigate causal inference in financial markets can now employ DML with much less complexity thanks to the automation and flexibility offered by the DoubleML library.

---

[2] https://www.imf.org/external/datamapper/fiscalrules/map/map.htm

[3] https://xgboost.readthedocs.io/en/stable/python/python_intro.html

## IV. EXPERIMENTAL ANALYSIS

### A. Analyzing the Models Predictive Performance

This study assesses the predictive power of gradient boosting and linear regression using the DML architecture in order to determine how well these methods work for financial return estimation. The accuracy of these underlying models is crucial for DML to extract meaningful treatment effects, hence the predictive performance of these models is critical. The study presents the terms "y-task" and "d-task" in this context to differentiate between the prediction of the treatment variable and the outcome variable, respectively as shown in Table II. The intricacies and computing requirements of financial data make it difficult to achieve strong prediction performance. Applications in policy-making or asset management may be hampered by inadequate predictive power since it can introduce substantial noise and produce estimates of the average treatment effect that may be deceptive. The findings demonstrate the intricacy of the connections within financial data by showing that gradient boosting has a superior predictive capacity than linear regression. Due to its intricacy, linear regression is unable to adequately represent the complicated dynamics of fund returns, so advanced modelling techniques are required. The results of the study are consistent with a wider range of literature in finance, which contends that fund performance is highly influenced by macroeconomic factors. These findings confirm that complex algorithms are essential for DML, particularly when dealing with time-dependent financial data. The DML framework's methodological approach involved adding delays to every variable in order to accommodate the time series character of financial data. This adaption illustrates the versatility of DML with time-dependent data, without compromising the framework's capabilities. The study's findings, which demonstrate the gradient boosting model's higher predictive ability, most likely come from combining ML and autoregressive techniques. The former provides the time dynamics, whilst the latter provides the ability to interpret high-dimensional, intricate functions. The training process and predicted performance of the ML model were improved by the substantial expansion of the observation set provided by panel data conversion. The success of the DML framework depended on the independence of observations, which was guaranteed by this transformation in conjunction with the addition of fixed effects. In comparison to an aggregated or individual fund method, there is a significant increase in data points, which highlights the significance of huge datasets for obtaining dependable ML results. But the research also raises the possibility of overfitting, particularly in the d-task models, which show high $R^2$ values for both fund kinds. Even after detrending, there is still a strong correlation between several variables and the treatment variable, indicating that there may be fundamental difficulties with the model even with the strict parameter tuning used to reduce overfitting. The contemporary monetary theory concepts that direct FRS to modify interest rates in response to the economic variables incorporated in the model may be the cause of this significant correlation. The outcomes support the use of gradient boosting in DML to analyse financial markets by demonstrating its ability to handle such difficult tasks. To maximise DML's use in finance and provide reliable and useful insights for financial sector stakeholders, however, the subtleties of financial data—such as temporal dependency and possible overfitting—need careful thought and additional study.

TABLE II
R² RESULTS

|  | Linear Regression | | Gradient Boosting | |
| --- | --- | --- | --- | --- |
|  | y-task | d-task | y-task | d-task |
| **R² (passive)** | 0.19 | 0.60 | 0.81 | 0.99 |
| **R² (active)** | 0.13 | 0.54 | 0.73 | 0.99 |

### B. Analyzing the Double Machine Learning Models

Using DML models, the study looks at how the increase rate of the interest rate affects fund returns for both actively and passively managed funds (refer to Tables III, IV, and V). While gradient boosting reveals this effect mainly in actively managed funds, linear regression models show a substantial negative causal effect on both fund types. It is important to comprehend the impact of a 1% increase in interest rates, which results in a significant drop in fund returns, especially for actively managed funds, when interpreting these findings. The predictive accuracy of the underlying models in the DML framework determines its usefulness in financial analysis. The outcomes highlight a nuanced relationship in the financial data that gradient boosting better explains than linear regression. This complexity is consistent with financial theories such as the CAPM, which suggests that macroeconomic fluctuations have a substantial effect on fund returns and that actively managed funds may outperform other funds in recessionary times. The substantial negative impact on actively managed funds supports conventional macroeconomic theories, and the study's findings cast doubt on the body of research on how funds perform in response to macroeconomic factors. But the absence of a meaningful correlation with passively managed funds raises questions and points to the need for more research, maybe because "passive" management can mean very different things to different people. The outcomes, especially for funds that are passively managed, may have been affected by challenges encountered during the data collecting and classification process. It's possible that the particulars of passive funds, including their investing goals, and the level of detail in the data reduced the apparent impact of interest rate changes. Furthermore, the examination indicated that the DML models might have been overfitted, which calls for a cautious interpretation of the noteworthy adverse effects observed in actively managed funds. Additional insights were obtained from the study's examination of the residuals from the DML models. The absence of patterns in residual plots (refer to Fig. 3) that would have suggested nonlinearity or heteroskedasticity in the models suggests their robustness. But there's room for improvement, especially when it comes to forecasting times of rate stability, as seen by the existence of outliers and the concentration of errors around zero interest rate movements. Given the volume of data and the inherent complexity of projecting passive fund returns, it is possible to explain the difference in predictive ability between active and passive funds, as demonstrated by the significant results for active funds and non-significant results for passive ones. This intricacy necessitates a more in-depth investigation, potentially because of the different nature of passive funds and how they react to changes in the macroeconomic environment.

---

[4] https://docs.doubleml.org/

TABLE III
RESULTS OF THE PASSIVELY MANAGED FUNDS' VIA DML

| Model | Coef. | SE | t | P > \|t\| | 2.5% | 97.5% |
|---|---|---|---|---|---|---|
| Linear Regression | -0.025 | 0.001 | -23.226 | 2.289e-119 | -0.027 | -0.023 |
| Gradient Boosting | 0.229 | 0.221 | 1.034 | 0.301 | -0.205 | 0.663 |

TABLE IV
RESULTS OF THE ACTIVELY MANAGED FUNDS' VIA DML

| Model | Coef. | SE | t | P > \|t\| | 2.5% | 97.5% |
|---|---|---|---|---|---|---|
| Linear Regression | -0.019 | 0.001 | -32.671 | 3.989e-2 | -0.021 | -0.018 |
| Gradient Boosting | -11.97 | 2.522 | -4.747 | 2e-6 | -16.91 | -7.028 |

TABLE V
DML ADJUSTED COEFFICIENTS THAT WERE CONVERTED TO ACCOUNT FOR A 1% (RATHER THAN A 100%) CHANGE

| Model | Passive | Active |
|---|---|---|
| Linear Regression | -0.00025 | -0.00019 |
| Gradient Boosting | 0.00229 | -0.1197 |

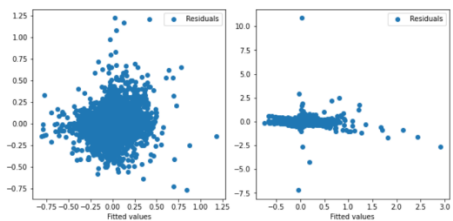

Fig. 3. Residuals vs fitted values

## V. RESULTS ANALYSIS

Three of the four models in the study's analysis of DML for determining causal impacts on fund returns produce significant outcomes, indicating the promise of DML for further research. Nevertheless, there are difficulties in predicting fund returns, which calls for intricate models and cautious result interpretation. The line separating actively managed funds from passively managed funds is becoming more hazy, which makes analysis more difficult and can necessitate the addition of more predictors to improve the models. One major problem is the computing intensity of DML, which is determined by the ML methods selected. The results of the study, in particular the treatment effects shown in the gradient boosting models, should be carefully examined because they might contain errors or accurately reflect the dynamics of intricate markets. The research was constrained by issues with data quantity and quality, with variations in data accessibility across active and passive funds potentially distorting findings. The homogeneity of treatment effects may have been impacted by the division of funds into active and passive categories, which may have ignored subgroups within these categories. It's possible that early data filtering and the conversion of quarterly macroeconomic variables into monthly data produced noise and removed essential information, respectively. The models' high computing requirements restricted the possibilities for parameter adjustment and the addition of new predictors, which possibly improved the study. Other financial indicators could potentially be incorporated into the study's methodology, which mainly uses macroeconomic factors, to create a more complete prediction model. To improve our understanding of causal effects in finance, future research should build on these findings by taking into account a wider range of funds, more diversified variables, and different ML approaches. Deeper insights into investment strategies and policy consequences could be obtained by validating DML's application in finance by extending the analysis beyond the U.S. market and investigating the predictive performance of different learners. In order to strengthen DML's credibility in financial causal inference and maybe increase its acceptance in business and management decision-making, future research should evaluate the underlying assumptions of the methodology.

## VI. CONCLUSION AND FUTURE WORKS

This study assesses the suitability of DML for measuring causal impacts on fund returns in the finance industry. DML can be used in finance with little modification, despite its computational hurdles, particularly when dealing with time series data. Although the reliability of these results need more confirmation through comparison with other ML models, the noteworthy results from the active funds dataset show the promise of DML. Despite being large, the estimated average treatment effect of interest rate fluctuations on fund returns should be interpreted cautiously because there may be other affecting factors. This study provides opportunities to combine ML and causal inference in finance, improving our understanding of market dynamics and helping stakeholders, including investors and policymakers, make well-informed decisions. Further study is required to examine the entire range of effects of macroeconomic factors on fund returns, strengthening the connection between scholarly research and real-world financial analysis.

## VII. DECLARATIONS

*A. Funding:* No funds, grants, or other support was received.

*B. Conflict of Interest:* The authors declare that they have no known competing financial interests or personal relationships that could have appeared to influence the work reported in this paper.

*C. Data Availability:* Data will be made on reasonable request.

*D. Code Availability:* Code will be made on reasonable request.

## REFERENCES

[1] M. C. Knaus, "Double machine learning-based programme evaluation under unconfoundedness," *Econometrics Journal*, vol. 25, no. 3, pp. 602–627, Sep. 2022, doi: 10.1093/ectj/utac015.

[2] J. C. Yang, H. C. Chuang, and C. M. Kuan, "Double machine learning with gradient boosting and its application to the Big N audit quality effect," *Journal of Econometrics*, vol. 216, no. 1, pp. 268–283, May 2020, doi: 10.1016/j.jeconom.2020.01.018.

[3] N. C. Chang, "Double/debiased machine learning for difference-in-differences models," *Econometrics Journal*, vol. 23, no. 2, pp. 177–191, May 2020, doi: 10.1093/ectj/utaa001.

[4] S. Aren and S. D. Aydemir, "The Factors Influencing Given Investment Choices of Individuals," *Procedia - Social and Behavioral Sciences*, vol. 210, pp. 126–135, Dec. 2015, doi: 10.1016/j.sbspro.2015.11.351.

[5] S. N. K. Appiah-Kubi *et al.*, "Impact of tax incentives on foreign direct investment: Evidence from africa," *Sustainability (Switzerland)*, vol. 13, no. 15, p. 8661, Aug. 2021, doi: 10.3390/su13158661.

[6] P. Cui *et al.*, "Causal Inference Meets Machine Learning," in *Proceedings of the ACM SIGKDD International Conference on Knowledge Discovery and Data Mining*, Aug. 2020, pp. 3527–3528. doi: 10.1145/3394486.3406460.

[7] S. H. Lin and M. A. Ikram, "On the relationship of machine learning with causal inference," *European Journal of Epidemiology*, vol. 35, no. 2. Springer, pp. 183–185, Feb. 01, 2020. doi: 10.1007/s10654-019-00564-9.

[8] H. Habib, G. S. Kashyap, N. Tabassum, and T. Nafis, "Stock Price Prediction Using Artificial Intelligence Based on LSTM– Deep Learning Model," in *Artificial Intelligence & Blockchain in Cyber Physical Systems: Technologies & Applications*, CRC Press, 2023, pp. 93–99. doi: 10.1201/9781003190301-6.

[9] J. Wang, T. Sun, B. Liu, Y. Cao, and D. Wang, "Financial Markets Prediction with Deep Learning," in *Proceedings - 17th IEEE International Conference on Machine Learning and Applications, ICMLA 2018*, Jan. 2019, pp. 97–104. doi: 10.1109/ICMLA.2018.00022.

[10] F. O. A. J. Mascarenhas, S.J., "The Turbulent Market of Modern Debt-overleveraged and Promoter-dominated Corporations," in *Corporate Ethics for Turbulent Markets*, Emerald Publishing Limited, 2018, pp. 189–213. doi: 10.1108/978-1-78756-187-820181007.

[11] J. Pearl, M. Glymour, and N. P. Jewell, *Causal Inference in Statistics A PRIMER*, vol. 4, no. 3. John Wiley & Sons, Incorporated, 2016. Accessed: Mar. 10, 2024. [Online]. Available: http://marefateadyan.nashriyat.ir/node/150

[12] N. Kamuni, S. Chintala, N. Kunchakuri, J. Narasimharaju and V. Kumar, "Advancing Audio Fingerprinting Accuracy with AI and ML: Addressing Background Noise and Distortion Challenges," in 2024 IEEE 18th International Conference on Semantic Computing (ICSC), Laguna Hills, CA, USA, 2024 pp. 341-345. doi: 10.1109/ICSC59802.2024.00064

[13] N. Kamuni, H. Shah, S. Chintala, N. Kunchakuri and S. Alla, "Enhancing End-to-End Multi-Task Dialogue Systems: A Study on Intrinsic Motivation Reinforcement Learning Algorithms for Improved Training and Adaptability," 2024 IEEE 18th International Conference on Semantic Computing (ICSC), Laguna Hills, CA, USA, 2024, pp. 335-340, doi: 10.1109/ICSC59802.2024.00063

[14] G. S. Kashyap, A. E. I. Brownlee, O. C. Phukan, K. Malik, and S. Wazir, "Roulette-Wheel Selection-Based PSO Algorithm for Solving the Vehicle Routing Problem with Time Windows," Jun. 2023, Accessed: Jul. 04, 2023. [Online]. Available: https://arxiv.org/abs/2306.02308v1

[15] G. S. Kashyap, A. Siddiqui, R. Siddiqui, K. Malik, S. Wazir, and A. E. I. Brownlee, "Prediction of Suicidal Risk Using Machine Learning Models." Dec. 25, 2021. Accessed: Feb. 04, 2024. [Online]. Available: https://papers.ssrn.com/abstract=4709789

[16] G. S. Kashyap, K. Malik, S. Wazir, and R. Khan, "Using Machine Learning to Quantify the Multimedia Risk Due to Fuzzing," *Multimedia Tools and Applications*, vol. 81, no. 25, pp. 36685–36698, Oct. 2022, doi: 10.1007/s11042-021-11558-9.

[17] S. Wazir, G. S. Kashyap, K. Malik, and A. E. I. Brownlee, "Predicting the Infection Level of COVID-19 Virus Using Normal Distribution-Based Approximation Model and PSO," Springer, Cham, 2023, pp. 75–91. doi: 10.1007/978-3-031-33183-1_5.

[18] G. S. Kashyap *et al.*, "Detection of a facemask in real-time using deep learning methods: Prevention of Covid 19," Jan. 2024, Accessed: Feb. 04, 2024. [Online]. Available: https://arxiv.org/abs/2401.15675v1

[19] N. Marwah, V. K. Singh, G. S. Kashyap, and S. Wazir, "An analysis of the robustness of UAV agriculture field coverage using multi-agent reinforcement learning," *International Journal of Information Technology (Singapore)*, vol. 15, no. 4, pp. 2317–2327, May 2023, doi: 10.1007/s41870-023-01264-0.

[20] G. S. Kashyap, D. Mahajan, O. C. Phukan, A. Kumar, A. E. I. Brownlee, and J. Gao, "From Simulations to Reality: Enhancing Multi-Robot Exploration for Urban Search and Rescue," Nov. 2023, Accessed: Dec. 03, 2023. [Online]. Available: https://arxiv.org/abs/2311.16958v1

[21] S. Wazir, G. S. Kashyap, and P. Saxena, "MLOps: A Review," Aug. 2023, Accessed: Sep. 16, 2023. [Online]. Available: https://arxiv.org/abs/2308.10908v1

[22] M. Kanojia, P. Kamani, G. S. Kashyap, S. Naz, S. Wazir, and A. Chauhan, "Alternative Agriculture Land-Use Transformation Pathways by Partial-Equilibrium Agricultural Sector Model: A Mathematical Approach," Aug. 2023, Accessed: Sep. 16, 2023. [Online]. Available: https://arxiv.org/abs/2308.11632v1

[23] S. Naz and G. S. Kashyap, "Enhancing the predictive capability of a mathematical model for pseudomonas aeruginosa through artificial neural networks," *International Journal of Information Technology 2024*, pp. 1–10, Feb. 2024, doi: 10.1007/S41870-023-01721-W.

[24] D. G. McMillan, "Which Variables Predict and Forecast Stock Market Returns?," *SSRN Electronic Journal*, Jun. 2017, doi: 10.2139/ssrn.2801670.